\documentclass[useAMS,usenatbib]{mn2e}

\usepackage{graphicx,epsfig}
\usepackage{ulem}

\def\be{\begin{equation}}
\def\ee{\end{equation}}
\def\ba{\begin{eqnarray}}
\def\ea{\end{eqnarray}}

\def\msun{M_\odot}

\def\ltsima{$\; \buildrel < \over \sim \;$}
\def\simlt{\lower.5ex\hbox{\ltsima}}
\def\gtsima{$\; \buildrel > \over \sim \;$}
\def\simgt{\lower.5ex\hbox{\gtsima}}
\def\etal{{et al.\ }}

\title[Gas outflows by disc-wide SNe]
{Launching of hot gas outflow by disc-wide supernova explosions}
\author[E. O. Vasiliev \etal]
       {Evgenii O. Vasiliev$^{1,2,3}$\thanks{E-mail:eugstar@mail.ru},
        Yuri A. Shchekinov$^{3,4}$, 
        Biman B. Nath$^{4}$\\
$^1$Southern Federal University, Stachki Ave. 194, Rostov-on-Don, 344090 Russia\\
$^2$Special Astrophysical Observatory of Russian Academy of Sciences, Nizhnii Arkhyz, Karachaevo-Cherkesskaya Republic, 369167 Russia \\
$^3$Lebedev Physical Institute of Russian Academy of Sciences, 53 Leninskiy Ave., 119991, Moscow\\
$^4$Raman Research Institute, Sadashiva Nagar, Bangalore 560080, India\\
}
\begin{document}
\date{Accepted 2018 December 32.
      Received 2018 December 31;
      in original form 2018 December 31}
\pagerange{\pageref{firstpage}--\pageref{lastpage}}
\pubyear{2119}
\maketitle

\label{firstpage}

\begin{abstract}
Galactic gas outflows are driven by stellar feedback with dominant contribution from supernovae (SN) explosions. The question of whether the energy deposited by SNe initiates a large scale outflow or gas circulation on smaller scales -- between discs and intermediate haloes, depends on SN rate and their distribution in space and time. We consider here gas circulation by disc-wide unclustered SNe with galactic  star formation rate in the range from {$\simeq 6\times 10^{-4}$ to $\simeq 6\times 10^{-2}~M_\odot$~yr$^{-1}$~kpc$^{-2}$, corresponding to mid-to-high  star formation observed in galaxies. We show that such disc-wide SN explosion regime can form circulation of warm ($T\sim 10^4$ K) and cold ($T<10^3$ K) phases within a few gas scale heights, and elevation of hot {($T>10^5$ K)} gas at higher ($z>1$ kpc) heights. We found that the  threshold energy input rate for hot gas outflows with disc-wide supernovae explosions is estimated to be of the order $\sim 4\times 10^{-4}$~erg~s$^{-1}$~cm$^{-2}$. We discuss the observational manifestations of such phenomena in optical and X-ray bands. In particular, we found that for face-on galaxies with SF ($\Sigma_{_{\rm SF}}>0.02~M_\odot$~yr$^{-1}$~kpc$^{-2}$), the line profiles of ions typical for warm gas show a  double-peak shape, corresponding to out-of-plane outflows. In the X-ray bands, galaxies with high SF rates ($\Sigma_{_{\rm SF}}>0.006~M_\odot$~yr$^{-1}$~kpc$^{-2}$) can be bright, with a smooth surface brightness in low-energy bands ($0.1\hbox{--}0.3$~keV) and patchy at higher energies ($1.6\hbox{--}8.3$~keV).}      
\end{abstract}



\section{Introduction}

\noindent

SNe explosions are thought to be the principal energy source for supplying heavy elements (metals) into galactic haloes and circumgalactic gas (CGM) 
by means of gaseous outflows {perpendicular to galactic discs} \citep{suchkov94,ferrara99,aditi18}. Two different scenarios are 
usually considered: {\it i)} the first suggests a starburst followed by multiple SN explosions in galactic centre, and {\it ii)} the second is connected with star formation in isolated OB-associations spread throughout the disc. The second is apparently more efficient in the sense that for a given total mechanical luminosity the amount of hot (X-ray) gas above the plane is larger \citep[see for recent discussion in][]{aditi18}. The differences between these scenarios, {their observational manifestations and large scale feedback have been widely debated since \cite{thomp05}, for more recent discussion see \citet{rub14,wib18,girichidis16,gatto17,field17,field18}.}  

In the discussion of mass transport by star formation (SF) activity, an issue of primary importance is the critical SF rate (SFR) able to 
to produce sufficient amount of energy for elevating mass from discs into haloes. In terms of 
the surface energy injection rate observational estimates for strong galactic winds driven by a central starburst, {the injection threshold rate has been estimated as} $\epsilon\sim 10^{-2}$ erg cm$^{-2}$ s$^{-1}$ \citep{lehnert96,heck02,heck03}, 
while for disc-wide case the estimates are an order of magnitude lower $10^{-3}$ erg cm$^{-2}$ s$^{-1}$ 
as inferred from synchrotron \citep{dahl95,dahl06} and from X-ray  and far-infrared emission \citep{tuel} of the 
haloes of edge-on galaxies. However, galactic winds do exist at a much lower energy injection rate: \citet{Li13} have reported 
observation of an X-ray bright galactic coronae with underlying SFR equivalent to  
only $\epsilon\sim 10^{-5}$ erg cm$^{-2}$ s$^{-1}$ at the lower end. {More recently, \cite{rub14} based on the analysis of 105 galaxies in the local universe ($0.3<z<1.4$) claimed ``no evidence for a threshold'' of the injection rate.} 

One can think that {the threshold energy rate can vary} depending on deposition rate of energy from SF and the regime in which this energy is deposited -- namely, whether SF occurs  mostly in large stellar clusters or in smaller clusters with only a few O-stars -- SN progenitors. In first case with large stellar clusters SN explosions can drive large scale outflows, in the second their feedback can {maintain} only circulation of gas within a few vertical scales. Observationally the former can give rise to such events as a {powerful} wind in M82, while the latter can reveal as a ``boiling'' galaxy NGC 253 \citep{sofue94} and massive intermediate haloes in edge-on galaxies like NGC 891 \citep{rossa04}. We focus here on the latter case with disc-wide unclustered SN explosions. 
 
Therefore, the lower limit of the energy injection rate -- if it exists, 
and physical processes that can cause this limit are of great importance for understanding global circulation of matter and heavy 
elements in the Universe. 

Currently most of the numerical experiments in driving disc-to-haloes outflows are based on the galactic wind model 
with continuous energy injection into a fixed volume, as {described} in the seminal paper by \cite{cc85}. Essentially it suggests that the energy 
injection rate is $\dot E\sim \nu_{\rm sn}E_{sn}$, with $\nu_{\rm sn}$ [s$^{-1}$] {and $E_{sn}$} being the rate of {SN} explosions in a given (injection) volume {and the explosion energy, correspondingly.}  
This assumption has been substantiated numerically in a spherically symmetric model for a uniform medium in \citep{sharoy14}: for a sufficiently 
high SN rate in a given volume hydrodynamic fields (velocity, density, pressure radial profiles) and the integral behaviour converge to the 
analytic self-similar solution described in \citep{cc85}. Moreover, in this limit the heating efficiency -- the fraction of SN energy converted 
into thermal and kinetic energy of the expanding (super)bubble, increases and asymptotically tends to $\eta_{\rm h}\simeq 0.4$ weakly depending on  
ambient density $n$ \citep{sharoy14,vns15}. For a low SN rate -- $\nu_{\rm sn}<10^{-12}$ yr$^{-1}$ pc$^{-3}$, the efficiency falls down 
$\eta_{\rm h}<0.08$, with a scaling $\eta_{\rm sh}\propto \nu_{\rm sn}^{0.2}n^{-0.6}$ \citep{vns15}. At such conditions, the launch and dynamics of 
outflows driven by stellar feedback can be described numerically in the luminosity driven mode with continuous mechanical luminosity   
$L\approx \nu_{\rm sn}\eta_{\rm h}E_{sn}$. 

However, this conclusion changes for supernova explosions in a vertically stratified ISM. In this case after a few SNe explosions the growing bubble 
can break through the disc and form a cavity growing upward, and all subsequent explosions expand in a very low density bubble interior with 
a minimum energy loss with $\eta_{\rm h}\sim 0.6$ \citep{yus18,field18}. This means that {an adequate description of outflows 
should involve injection of energy from individual SN  
explosions.} Moreover, the approach based on a continuous liminosity driven model is inherently an average description, i.e. 
$L=\nu_{\rm sn}E_{sn}=E_{sn}\langle\sum\limits_{i}^{}\delta(t-t_i)\delta({\bf r-r}_i)\rangle$, where $t_i$ and ${\bf r}_i$ is the time 
and the position of an $i$-th SN explosion. As such it may miss dynamical and morphological details  
connnected with a random character of explosions. For separate individual explosions mechanical luminosity is a random function  
$L(t,{\bf r})=E\sum\limits_{i}^{}\delta(t-t_i)\delta({\bf r-r}_i)$, and its dynamical feedback depends on particular configuration 
of the sets [$t_i$] and [${\bf r}_i$] and their statistical properties, 
such as characteristic separation between in the set of $t_i$ and in the set of ${\bf r}_i$. 

The issue of mass transport by individual supernovae spread randomly in stellar discs has been address in several previous studies, including \citet{kim18}. However, the question we address in this paper is not only the efficiency of such energy injection, but the possible observational manifestations, and the observational diagnostics that can tell us about the mass transport rate. The paper is organized as follows. Section 2 contains model description and numerical setup, in Section 3 we describe the results, while in Section 4 we discuss {observational} manifestations, details of the model, Section 5 summarizes the results.


\section{Model description and numerical setup}\label{modl}

\noindent

We carry out 3-D hydrodynamic simulations (Cartesian geometry) of multiple SNe explosions in the galactic disc. We set up a gaseous disc to be initially in the hydrostatic equilibrium in a gravitational potential {\citep[as in many previous papers, see e.g.,][]{avillez00,hill12,walsch15,li17}, which consists of two components: a dark matter (DM) halo and a baryonic disc. In cylindrical coordinates with $z$-axis perpendicular to the disc, the $z$-component of the gravitational acceleration due to the DM halo is calculated from a Navarro-Frenk-White profile: $g_{DM}(z) = GM_{DM}(r)z/r^3$, where $M_{DM}$ is the DM mass enclosed within radius $r$ { and we adopt virial radius of the halo equal to 200~kpc and concentration parameter $c=12$.} }

\begin{figure}
\center
\includegraphics[width=7cm]{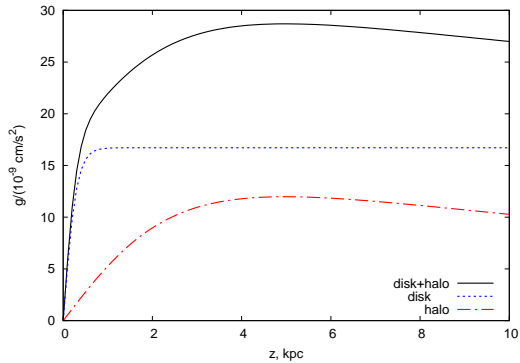}
\caption{
Gravitational acceleration vs height $z$ at {the cylindrical radius 3 kpc} as defined by Eq. (\ref{poten}). Dotted and dot-dashed lines show contributions from the stellar disc and from the halo, their sum is shown by solid line.   
}
\label{3kpcgr}
\end{figure}

{
The baryonic disc is assumed to be self-gravitating with an isothermal velocity dispersion. The acceleration perpendicular to the disc is $g_*(z) = 2\pi G \Sigma_* {\rm tanh} (z/z_*)$, where $\Sigma_*$ and $z_*$ are the stellar surface density and scale height of the stellar disc, respectively. Contribution from a gaseous disc is included, following \citet{li17}, by dividing $g_*(z)$ by $f_* = \Sigma_*/(\Sigma_* + \Sigma_{gas})$, that implicitly suggests the gaseous disc to be non-self-gravitating. Thus, the $z$-components of the total gravitational acceleration is calculated as \citep{li17}
\be\label{poten}
 g(z) = GM_{DM}(r){z\over r^3} + {1\over f_*}2\pi G \Sigma_* {\rm tanh} \left({z\over z_*}\right). 
\ee
Figure \ref{3kpcgr} shows $g(z)$ within the heights of interest at the cylindrical radial distance $R=3$ kpc, { for which the value of $z_*$ is assumed to be equal 0.3~kpc as it has been estimated in the solar neighbourhood \citep{gilmore83}, the stellar surface density equals to $\Sigma_* = 180~\msun$~pc$^{-2}$ and gas surface density $\Sigma_g = 3.6~\msun$~pc$^{-2}$. The} contributions from the stellar component and from the halo are shown by dotted and dot-dashed lines. Similar model for the potential gradient has been used for simulations of multiphase galactic outflows \citep[e.g.][]{li17}, and is consistent with the model of gravitational potential built by \cite{kalb03} at the cylindrical radial distance 3~kpc from the Milky Way centre \citep[see Figure 7 in][]{kalb03}. It is adopted as a fiducial case in our simulations, unless otherwise specified. }

The gas density profile is found from the hydrostatic equlibrium. At large heights the number density is kept uniform with at $10^{-3}$~cm$^{-3}$. The gas surface density $\Sigma_g = 3.6~\msun$~pc$^{-2}$ corresponds to the number density at the midplane 1~cm$^{-3}$. Initially the gas  temperature is $9\times 10^3$~K; the gas metallicity is kept constant equal to the solar value within the whole computational domain. Standard simulations are performed with a physical cell size of 2~pc in the computational domain with $256\times 256\times 768$ cells, that corresponds to $512\times 512\times 1536$~pc$^3$. Note that in the case of high SNe rate a significant part of gas from the disc may leave our standard computational box. In this case we expand the box from 1.5~kpc to 16~kpc, and increase the physical cell size to 4~pc. We ensured that this increase does not affect the evolution of the physical parameters considered here, e.g. velocity distribution, mass fraction, emission characteristics.

SNe are distributed randomly and uniformly in the galactic plane and exponentially in the vertical direction. The scale height is 1/3 of the stellar scale height, i.e. $h_{\rm SN}=0.1$~kpc. SN events follow one-after-one uniformly in time; we ran models with the averaged time delay $\Delta t$ from $10^6$ to $10^4$~yrs. Using the sizes of the computational domain and SN scale height, one can easily find the volume ($\nu_{SN}$) and surface ($\Sigma_{SN}$) SNe rates. Thus, the volume SNe rate $\nu_{\rm SN}$ ranges from $2\times 10^{-14}$ to $2\times 10^{-12}$~yr$^{-1}$~pc$^{-3}$. The surface density of star formation rate corresponding to the volume SNe rate equals $\Sigma_{_{\rm SF}} = 2h_{_{\rm SN}} \nu_{_{\rm SN}} m_0$, where $m_0 = 150~\msun$ for a Salpeter IMF, or in the normalized form
\ba\label{sigm}
 \Sigma_{_{\rm SF}} \simeq 5.7\times 10^{-2} {\rm \msun yr^{-1} \ kpc^{-2}} 
                                                         \left({ h_{_{\rm SN}} \over 0.1~{\rm kpc}} \right)  \times \\ \nonumber
                                                  \times \left({ \nu        \over 2\times 10^{-12} {\rm yr^{-1} pc^{-3}}} \right).
\ea
One should stress here that we do not fit directly the adopted SN (and correspondingly SF) rates to the Kennicutt-Schmidt (KS) law. Indeed, 
to the highest SN rate considered here the $\Sigma_{\rm SFR}$ value should correspond to the range of gas surface density $\sim 10-100~\msun$~pc$^{-2}$ in the KS relation \citep{schmidt59,kennicutt98} and is obviously much higher than that adopted in our simulations. Although our focus here is a particular aspect of stellar feedback -- gas circulation in galactic discs, it is worthnoting that the KS relation in the high surface density end is inferred from measurements of $H_\alpha$, HI, far-infrared continuum and CO emissions averaged over a relatively low angular resolution \citep[see discussion in][]{kennicutt98,big08}. As such it can be applicable only to large areas of galactic discs. At sub-kpc scales it may not be valid because star-forming molecular clouds are quickly (within a few Myr to tens of Myr) destroyed by feedback from massive stars, and the SNe explode mostly in regions of lower density (see for discussion of observational aspects in \citet{kennicutt-review12}, and theoretical arguments in { \citet{krum12,kv17,semenov18}}). From this point of view our models illustrate transport of mass from the regions of galactic discs cleaned from dense molecular gas by stellar feedback. It is obvious that SN driven ejection of denser molecular gas from discs with higher $\Sigma_g$ is less efficient. 

{It is also worth emphasizing that individual SNe are assumed here to be immersed in a smooth interstellar environment with the density distribution corresponding to an average density without considering effects from a parent molecular cloud. Such a description is common when large scale effects in s SN-driven ISM are concerned \citep[see discussion in][ and references therein]{vns15,vsn17,li17,li17a,aditi18,field18}. On the contrary, study of many issues connected with ``microscopic'' scales of SN ejecta interaction with dense molecular environment in the parent molecular clous, in particular, with generation of cosmic rays in superbubbles, do concern dynamics of a SN shock propagating in an immediate surrounding dense molecular material \citep[see review in][]{vikram}. In this latter case, only a fraction (still rather uncertain) of SN energy is deposited into ISM on large scales comparable to the scale height of gas. Therefore, the numerical values of SN rates assumed in numerical simulations of stellar feedback on large galactic scales has to be increased by an uncertain factor $f\sim 1$ in order to account supernova explosions hidden in their parent molecular clouds.   
}

Locations of the supernova rate do not correlate the gas density. In all models supernova rate does not change with  time; this is equivalent to a constant star formation rate. The random times and locations of supernovae are chosen at initialization, such that for all runs with a given configuration, supernovae explode in the same locations. Note that for very low SNe rates remnants of  individual SNe can cool down separately before they merge into a collective bubble. We restrict our simulations within a timescale of a single starburst $\simlt 30$~Myr. We inject the mass and energy of each SN in a region of radius $r_0=3$ pc, with the energy, $10^{51}$~erg, injected as thermal energy. 

The code is based on the unsplit total variation diminishing (TVD) approach that provides high-resolution capturing of shocks and prevents unphysical oscillations. We have implemented the Monotonic Upstream-Centered Scheme for Conservation Laws (MUSCL)-Hancock scheme and the Haarten-Lax-van Leer-Contact (HLLC) method \citep[see e.g.][]{toro99} as an approximate Riemann solver. This code has successfully passed the whole set of tests proposed in \citet{klingenberg07}. 

Simulations are run with radiative cooling processes with a tabulated non-equilibrium cooling function fitting the calculated one \citep{v11,v13}. The fitted function is obtained for  gas cooling isochorically from $10^8$~K down to 10~K. The non-equilibrium calculation \citep{v11,v13} includes kinetics of all ionization states of H, He, C, N, O, Ne, Mg, Si, Fe, as well as kinetics of molecular hydrogen at $T<10^4$~K. 

We apply a diffuse heating term representing the photoelectric heating of dust grains \citep{dust-heat}, which is thought to be the dominant heating mechanism in the interstellar medium. In our simulations, the heating rate is assumed to be time-independent and exponentially decreasing in the vertical direction with the scale height of the ISM disc. Such an assumption is sufficient to stabilize radiative cooling of ambient gas at $T=9\times 10^3$ K. Any deviation of the heating rate in the unperturbed gas violates the balance between cooling and heating and stimulates thermal instability, and leads to redistribution of gas mass in the interstellar disc \citep[see e.g. in][]{avillez00,hill12}. In order to avoid contaminations of such {effects we follow \cite{li17} to assume} the heating rate exponentially decreasing upwards across the whole computational domain. 

\begin{figure*}
\includegraphics[width=8.5cm]{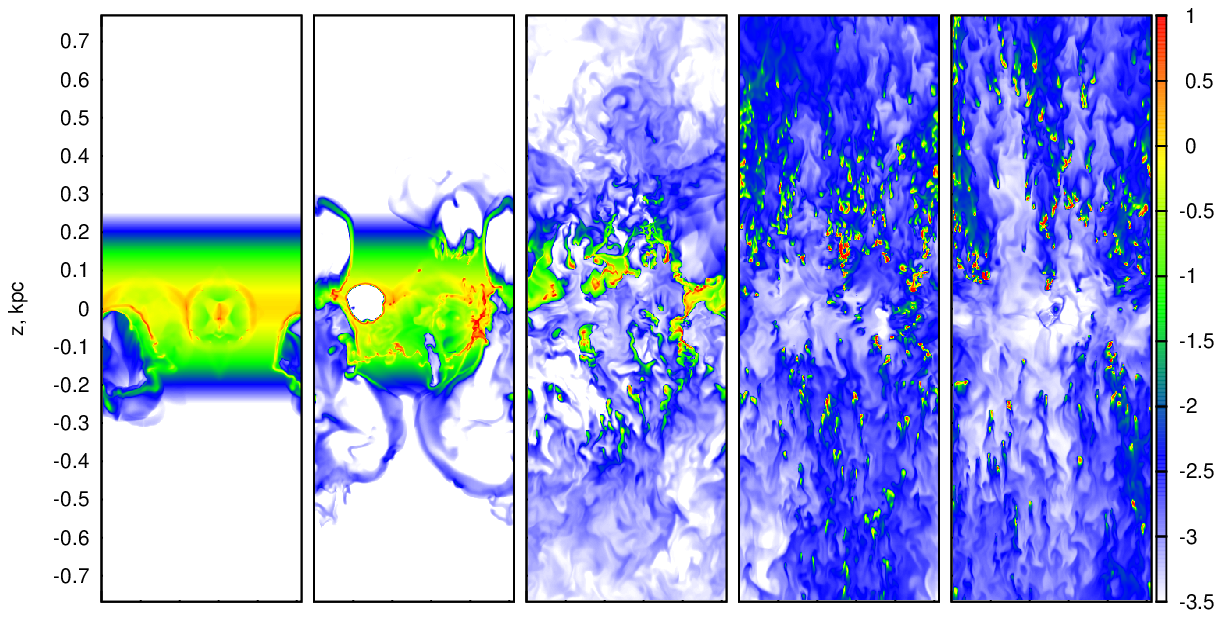}
\includegraphics[width=8.5cm]{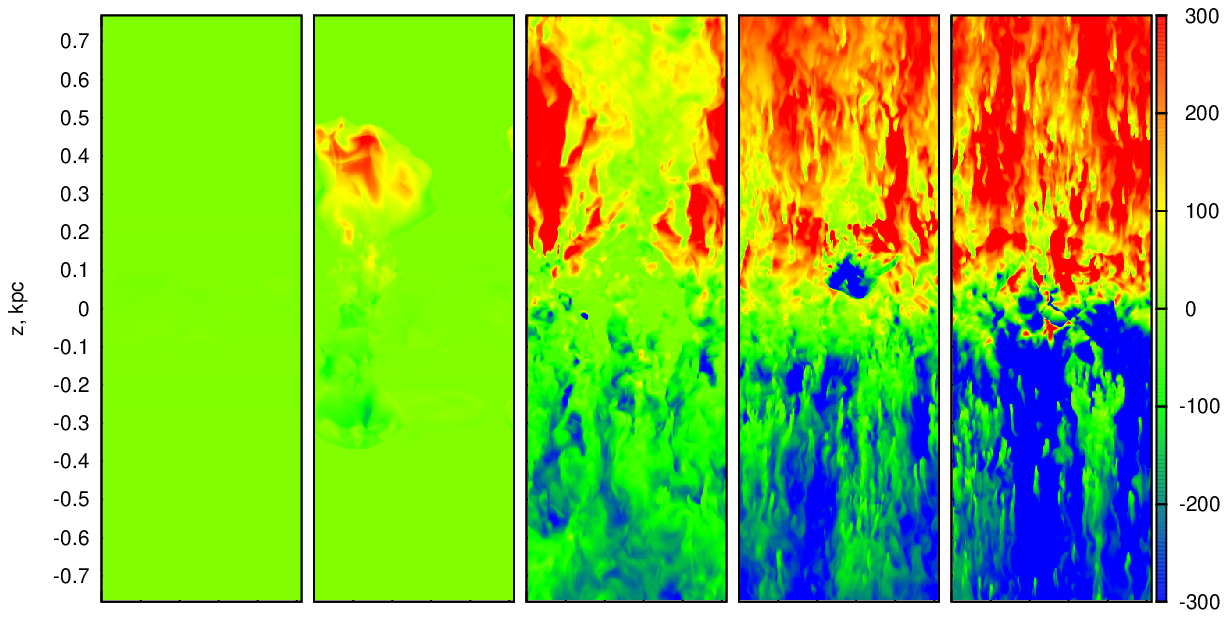} \break
\includegraphics[width=8.5cm]{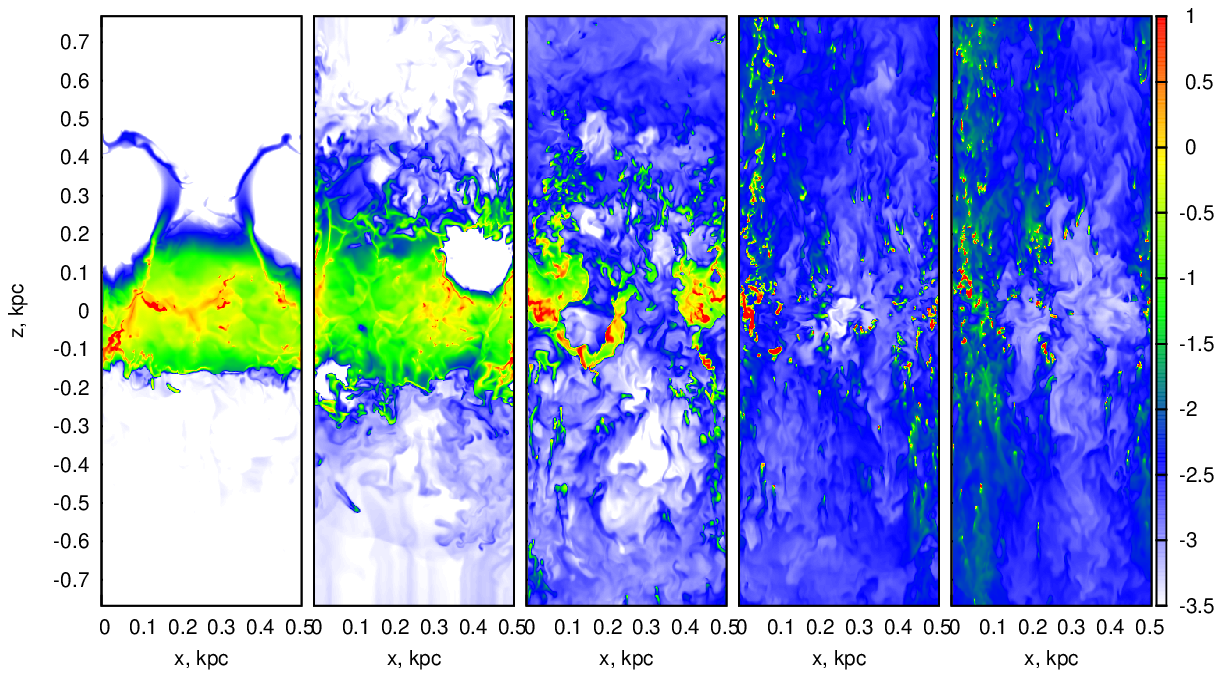}
\includegraphics[width=8.5cm]{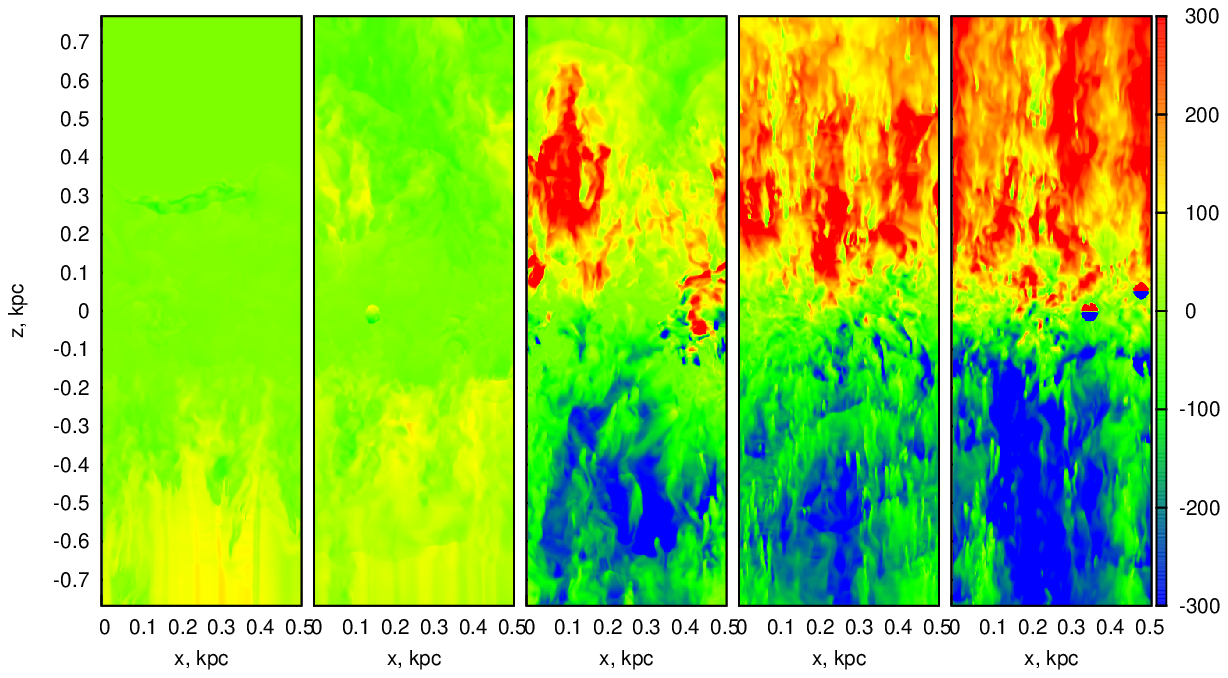}
\caption{
{2D slices showing {gas number} density ({log$(n$, cm$^{-3})$,} left panels) and velocity {in $z$-direction} ({$v$ in km/s,} right panels) distribution in outflows for different SN rate: 
~$2\times 10^{-14}$,~$6\times 10^{-14}$,~$2\times 10^{-13}$, ~$6\times 10^{-13}$, ~$2\times 10^{-12}$~yr$^{-1}$~pc$^{-3}$ -- left to right. 
Upper row of panels shows density and velocity fields at $t=10$ Myr, lower row of panels is for $t=30$ Myr.} 
}
\label{fig-1sn-por-10M}
\end{figure*}

Recently \citet{field17} studied the dynamics of an outflow generated by individual supernovae spread throughout the galactic disc in relatively small galaxies with radial scale of $R_d\sim 300$ pc. It follows from their consideration that among other factors a non-planar (spherical or quasi-spherical) geometry of the gravitational field and the ISM disc seems to be  an essential condition for launching a large scale galactic outflow. {When higher mass (disc) galaxies are concerned it is clear that certain conditions have to be fulfilled in order that the energy source -- SN explosions in our case, would be able to expel gas at heights where non-planar effects come into play. It occurs normally at scale heights comparable to the disc {radial scale \citep{kalb03}}.} However, our paper aims at a different aspect of the problem --- how supernovae can throw gas to small length scales (a few gas scale heights), corresponding to low to intermediate size gaseous halos, observed in Milky Way size edge-on galaxies \citep{dahl95,dahl06,heesen18}, where the effects of diverging quasi-spherical geometry are weaker than in dwarf galaxies. 


\section{Results}

The ability of a single SN to break through the ISM layer and eject matter and energy out of the disc into the extra-planar space, is determined by interrelation between the explosion energy $E$, the gas scale height $z_0$ and the in-plane density $\rho_0$. It can be 
loosely written as $\rho_0z_0^3\simlt\eta_{\rm h} Ec_s^{-2}$, where $\eta_{\rm h}\sim 0.1\hbox{--}0.3$ is the heating efficiency -- the fraction of 
energy left after radiative cooling, $c_s$, is the sound speed in ambient gas \citep[see, e.g., discussion in][]{ft00}. For typical parameter values it 
can be reduced to $nT_4<0.2\eta_{\rm h}z_{100}^{-3}$, which cannot be fulfilled for the ISM pressure $p\sim 3000\hbox{--}10^4$ K cm$^{-3}$ and for 
$z_0>50$ pc. For a SN placed at a height $z$ above the galactic plane this condition softens approximately by factor of $e^{-z/z_0}$, such that 
even a single SN explosion can break through the disc and eject mass and energy out of it. Then this means that if the scale height of supernovae $z_{\rm sn}\simgt z_0$, even 
a relatively weak SF is able to heat halo and supply into it gas from exploding SNe and partly from the ISM disc as well. It is important to note, that even though the fraction of such outstanding SNe seems to lie in the range of Poisson noise, the two circumstances amplify their contribution: one is because of the above mentioned weakened condition for breaking through the disc, the other is that for a SN exploding at $z\simgt z_0$ the remnant expands upward faster than for those exploding in the plane.
This circumstance can come into play in conditions with low SF and SN rates, when deviations from an average behavior may be as considerable 
as $\propto\nu^{-1/2}$. 


\begin{figure*}
\center
\includegraphics[width=4cm]{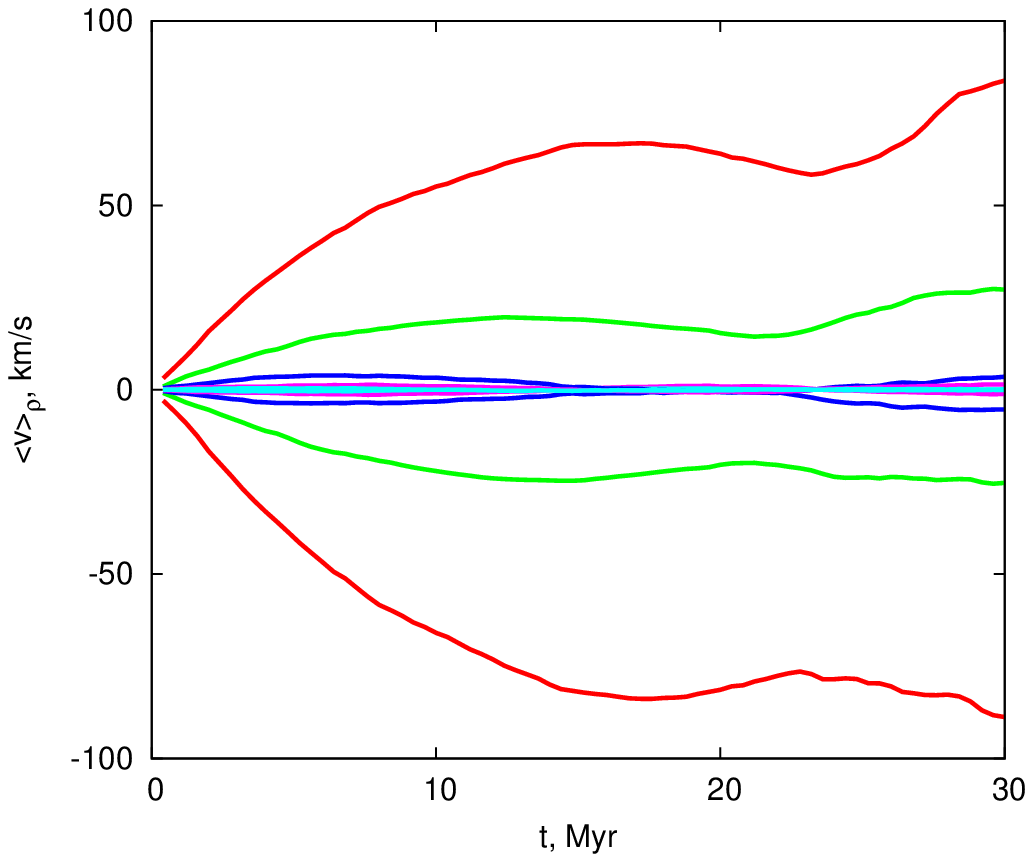}
\includegraphics[width=4cm]{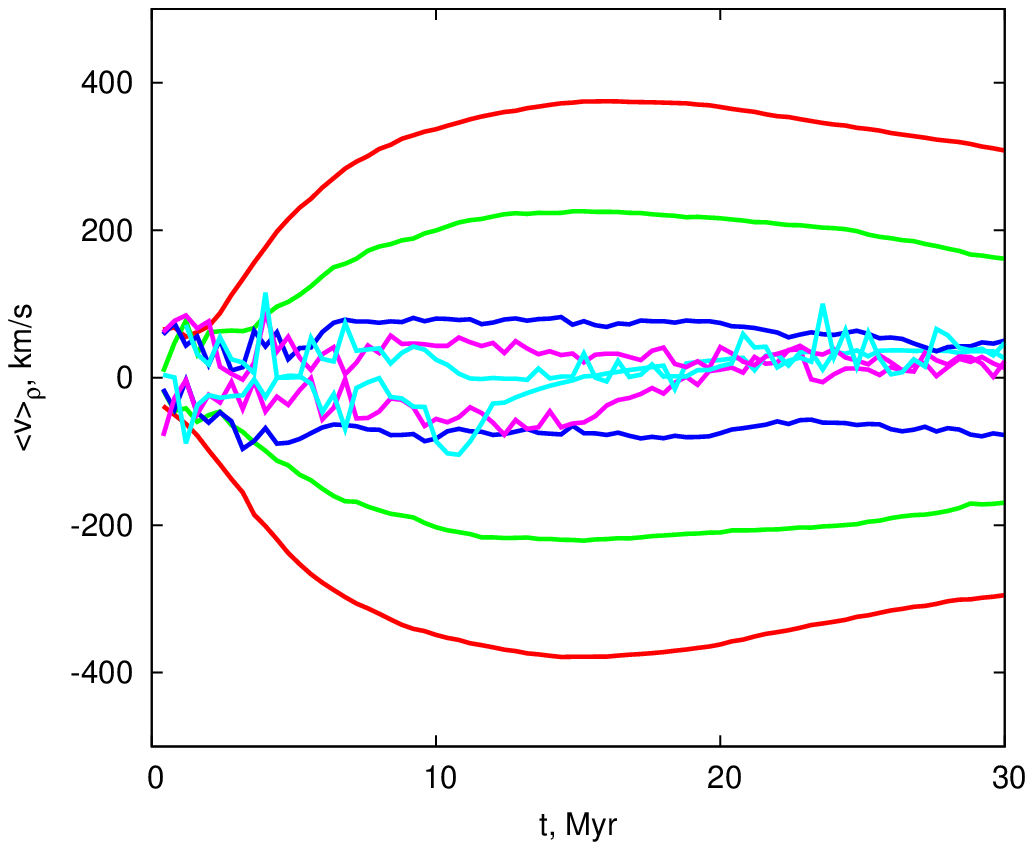} 
\includegraphics[width=4cm]{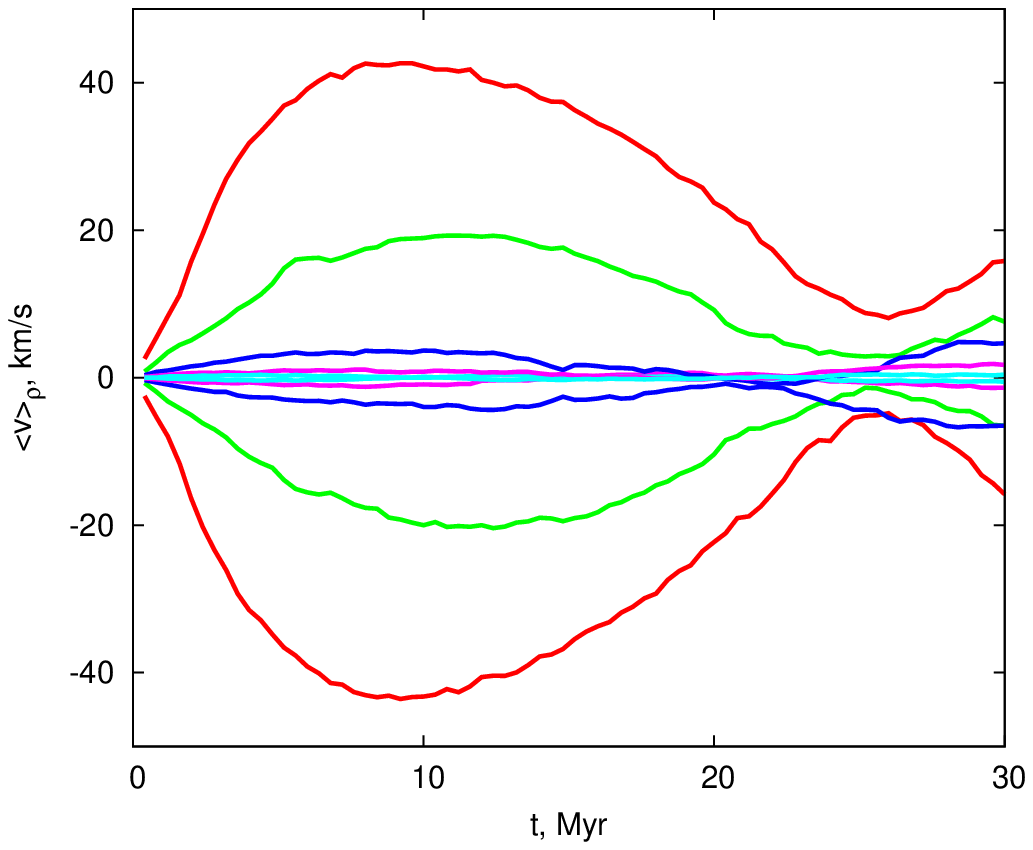}
\includegraphics[width=4cm]{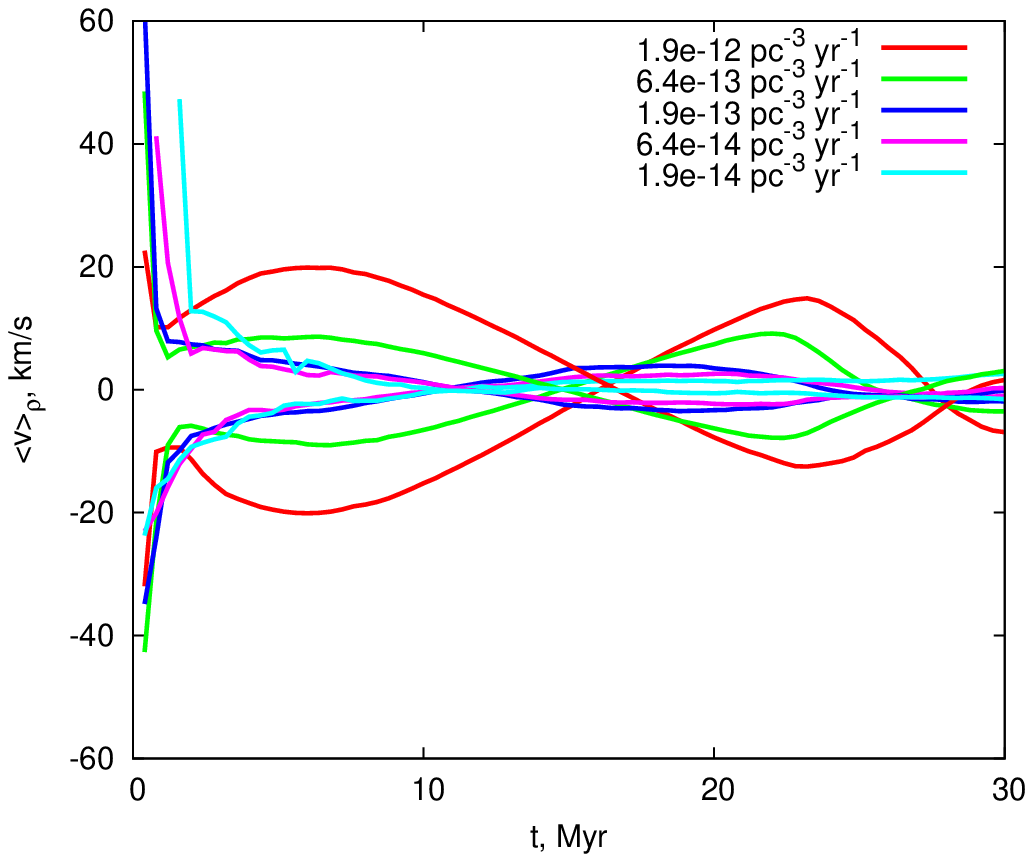}
\caption{{Mass-average velocity as a function of time for the models shown in Fig. \ref{fig-1sn-por-10M}: colors code SN rate from the lowest (cyan) to the higher (red) as shown in the legend; left-to-right panels show the mass-average velocity for the whole temperature range (left), for hot phases with $T>10^5$ K (second), for the warm phases with temperature $10^3<T\leq 10^5$ K (third), and for the cold phases with temperature $T\leq 10^3$ K  (right); quasi-periodical variations of velocity for low-temperature phase reflects nearly ballistic free-fall motions of dense clouds with the characteristic time $t_{\rm ff}\simeq 15$ Myr for clouds within $|z|\simlt 300$ pc in model with the highest SN rate.  }
}
\label{fig-v-mass}
\end{figure*}

Figure \ref{fig-1sn-por-10M} {(left column of panels)} {shows  slices representing density distribution in the computational zone at times $t=10$ Myr and $t=30$ Myr for different SN rates: ~$2\times 10^{-14}$,~$6\times 10^{-14}$,~$2\times 10^{-13}$, ~$6\times 10^{-13}$, ~$2\times 10^{-12}$~yr$^{-1}$~pc$^{-3}$ (the corresponding SF surface density rates range within $\Sigma_{_{\rm SF}}\simeq (6\times 10^{-4}\hbox{--}0.06)\, M_\odot$~yr$^{-1}$~kpc$^{-2}$, assuming the SN rate per mass $\nu_m\simeq (150~M_\odot)^{-1}$).} {The porosity factor $q=2\pi R_0^2 z_0\nu t_0\sim z_{0,100}\nu_{-13}n^{-1}$ increases from $\sim 0.2$ on the left to $\sim 2$ on the right; here $t_0=3t_r\sim n^{-1/3}$ Myr with $t_r$ being the characteristic time when an expanding remnant enters the radiative phase, $R_0$, the remnant radius at $t=t_0$, $z_{0,100}=z_0/100$ pc, $\nu_{-13}=\nu/10^{-13}$ in units yr$^{-1}$~pc$^{-3}$. Factor $q$ characterises the fraction of overlapped remnants from individual supernovae -- in the left side overlapping is small and remnants from individual supernovae remain isolated without the production of a collective effect.} It is clearly seen that for galactic discs with stellar scale height $z_\ast=100z_{_{\ast,100}}$ pc models with the surface SF rate smaller than $\Sigma_{\rm SF}<0.006~z_{_{\ast,100}}M_\odot$~yr$^{-1}$~kpc$^{-2}$ (SN rate $\nu<2\times 10^{-13}$~yr$^{-1}$~pc$^{-3}$) {for which $q<1$,} may show strong deviations from symmetry between hydrodynamic fields above and below the plane -- reminiscent  of spontaneous symmetry breaking.  

\begin{figure}
\center
\includegraphics[width=4cm]{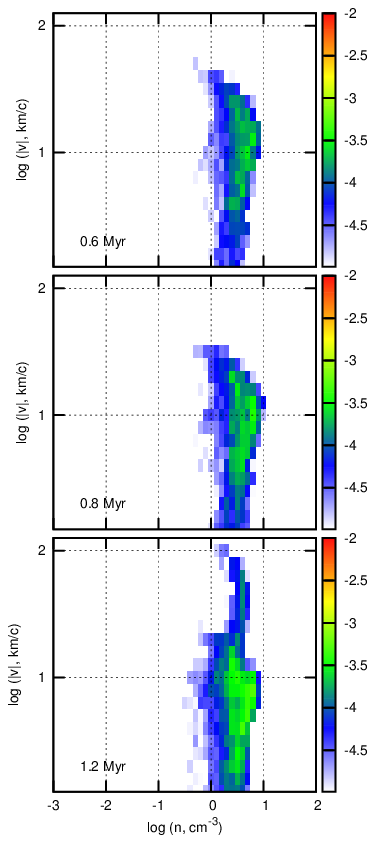}
\includegraphics[width=4cm]{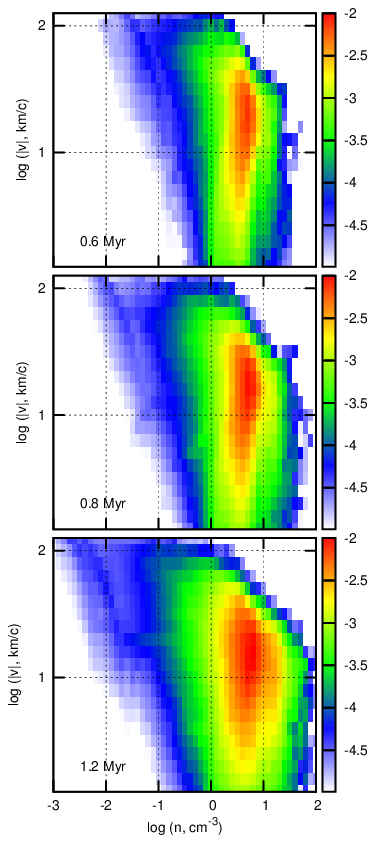} 
\caption{Velocity versus density within the initial $\sim 1$ Myr episode of SNe activity for the model  
with $\nu=2\times 10^{-14}$ yr$^{-1}$ pc$^{-3}$ (left) and $\nu=2\times 10^{-12}$ yr$^{-1}$ pc$^{-3}$ (right); times are 0.6, 0.8, 1.2 Myr from top to bottom. Color coding corresponds to the mass fraction of gas in a given range of $n$ and $|v|$. It is seen that after $0.6\hbox{--}0.8$ Myr, gas accumulates in relatively dense $0.3<n<10^2$ cm$^{-3}$ clumps.
}
\label{fig-v-n-1.5Myr}
\end{figure}

In general, the outflow weakens with time because a fraction of gas cools and condenses into colder and denser clouds and cloudlets under compression from overlapping shocks. When sufficient time has elapsed most massive of them turn to fall down even though SN rate stay constant. Clouds start forming along the outflow due to thermal instability enhanced by shock compression at times $t\sim 0.1\hbox{--}0.3$ Myr. They are seen on the density field in Figure~\ref{fig-1sn-por-10M} as small dense clumps. Once clouds are formed the interrelation between pressure and gravity forces acting on them changes: the dominant forces on the are ram pressure from hot diffuse ambient gas $\propto R_{\rm cl}^2$ and gravitational force $\propto R_{\rm cl}$, $R_{\rm cl}$ being cloud radius. This circumstance results in a gradual decrease of the mass-averaged gas velocity {between $t\simeq 15$ and $t\simeq 25$ Myr} as shown in Figure~\ref{fig-v-mass} {(first panel)}. It is also seen if one compares the velocity fields at $t=10$ Myr and $t=30$ Myr in Figure~\ref{fig-1sn-por-10M}. Models with smaller SN rate do not form a developed gas circulation, but rather a sporadic isolated growing or contracting bubbles without such a pronounced collective effect as seen in case of high SN rate. It is a consequence of the fact that the porosity factor for these models is small and there is no percolation of different remnants. In models with high SN rate the mass-averaged velocity is regulated by hot gas phase before $t\simlt 20$~Myr, even though its small mass (second panel). Quasi-periodic oscillations are clearly seen for the cold phase (right panel). It reflects nearly ballistic motions of dense clouds within $|z|\simlt 300$ pc with the characteristic free-fall time $t_{_{\rm ff}}\simeq 15$ Myr for the highest SN rate and decreasing to $t_{_{\rm ff}}\simeq 10$ Myr in for the lowest SN rate. Warm phases also seem to show quasi-periodic motions with longer characteristic times mediated by pressure (third panel). A small increase of the mass-average velocity for all phases (left panel) is due to such a quasi-periodic increase in the velocity of warm phases (third panel) at $t\sim 30$ Myr.

In order to illustrate details of the initial (within $t\sim 1$ Myr) development of outflows in the low and high SN rates we show in Figure~\ref{fig-v-n-1.5Myr} the `velocity-density' {distributions} for $2\times 10^{-14}$ (left panel{s}) and $2\times 10^{-12}$~yr$^{-1}$~pc$^{-3}$ (right panel{s}) SN rates. It illustrates that such a transition from the outflow in the early episodes toward circulation of gas and subsequent settling down occures within the initial 1.5 Myr: SN  shock waves accelerate diffuse gas, compress it and when the shock fronts decelerate below $v_s\simlt 100$ km s$^{-1}$ they {cool radiatively and form denser clumps via thermal instability}. Further fragmentation under thermal instability and formation of clumps occurs in this velocity range with subsequent slowing of clumps due to gravity and ram pressure. In Figure \ref{fig-v-n-1.5Myr} it manifests in a gradual growth of mass of a relatively dense ($0.3<n<10^2$ cm$^{-3}$) and low velocity ($|v|\simlt 70$~km~s$^{-1}$) gas after $t=0.6\hbox{--}0.8$ Myr -- the distribution function on Figure \ref{fig-v-n-1.5Myr} widens from the high density edge $n\simeq 20$ cm$^{-3}$ at $t=0.6$ Myr to the high density edge $n\simeq 50$ cm$^{-3}$ at $t=0.8$ Myr. Only hot low density gas continues flowing out. The amount of such gas at lower SN rates is negligible, and it remains such in further development as can be judged from Figure~\ref{fig-1sn-por-10M}. 

\begin{figure}
\center
\includegraphics[width=7cm]{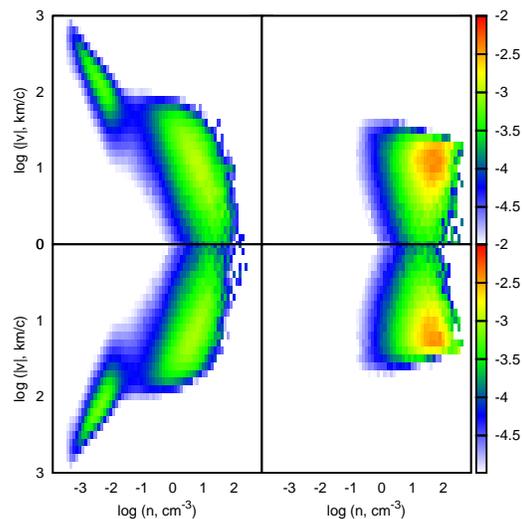}
\caption{Contribution of outflowing -- {\it left panel}, and infalling -- {\it right panel}, gas into the `velocity--density' diagram; upper panels are for gas above, and lower panels for gas below the plane {for SN} rate $\nu=2\times 10^{-12}$ yr$^{-1}$ pc$^{-3}$, at time $t=20$ Myr. Color coding is as in Fig. \ref{fig-v-n-1.5Myr}. It seems worth mentioning that denser gas falling on the galactic plane has velocities corresponding to the intermediate-velocity clouds (IVC) in the Milky Way galaxy \citep{wak01}. 
}
\label{fig-v-u-d-20Myr}
\end{figure}

A small fraction of hot low density gas penetrates between the denser clumps and expand with higher velocities $v>70$ km s$^{-1}$ upward, while denser gas confined in clumps and clouds decelerates below $v<70$ km s$^{-1}$ and eventually falls on to the disc at $t>10$ Myr under gravitational force. It is clearly seen in Figure \ref{fig-v-u-d-20Myr}, {where a similar interrelation between the velocities and densities as in Fig. \ref{fig-v-n-1.5Myr} are shown separately for positive and negative $z$ and outflowing and infalling velocities for the model with $\nu=2\times 10^{-12}$ yr$^{-1}$ pc$^{-3}$ at the time 20 Myr. In the left panel Fig. \ref{fig-v-u-d-20Myr} $(v,n)$-plane is shown for outflowing gas -- i.e. gas moving upwards with the upper and lower sub-panels showing $z>0$ and $z<0$ parts of the gaseous disc. The right-side panel shows interrelation between the velocities and densities of the gas moving downwards, i.e. corresponding to the inflow, and again the upper and lower parts of this panel show positive and negative $z$ of the disc. In other words, the contributions of outflowing and infalling gas into the `velocity--density' diagram separately for outflowing and infalling gas above and below the galactic plane at time 20 Myr are shown. It is seen that the dominant contribution to the domain of dense gas with velocities $|v|<100$ km s$^{-1}$ on both sides of the ISM disc comes from gas falling back onto plane. They look very symmetric and it exactly corresponds to quasi-periodic variations of the mass-average velocity for the cold $T<10^3$ K phase in Fig. \ref{fig-v-mass}.  }

{At longer time scales gaseous discs with low SF and SN rates -- lower than $\Sigma_{\rm SF}<0.006~z_{_{\ast,100}}M_\odot$~yr$^{-1}$~kpc$^{-2}$ and {$\nu< 2\times 10^{-13}$~yr$^{-1}$~pc$^{-3}$}, correspondingly, slightly puff out and remain apparently confined within the same scale height (see discussion below).} 

\begin{figure*}
\includegraphics[width=14cm]{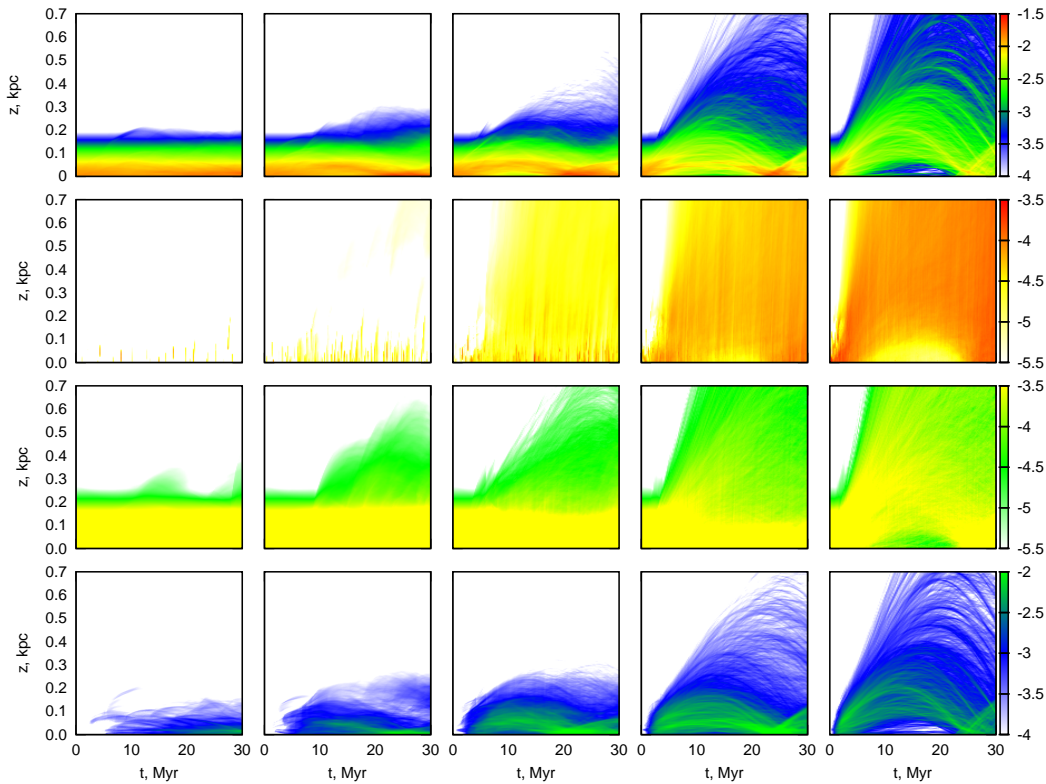}
\caption{Fraction of gas mass at a given height $z$ versus time for the models depicted on Fig. \ref{fig-1sn-por-10M}. {\it Top to bottom}: all 
gas, hot -- $T\geq 10^5$ K, warm -- $10^3<T<10^5$ K and cold -- $T\leq 10^3$ K components. {Color bars in the right show the mass fraction accumulated in a given gas component. }
}
\label{fig-mass-t}
\end{figure*} 

Therefore, this process -- gas clumping due to fragmentation, results in the decrease of mass outflow rate on longer times $t\simeq 15\hbox{--}20$ Myr, even when SN explosions are active. {Figure \ref{fig-mass-t} displays the dependence of gas mass fractions $f_m(z)$, $f_{m,h}$ and $f_{m,c}$  deposited at a given height $z$ on time for the models shown in Fig. \ref{fig-1sn-por-10M}. The upper row shows the fraction of gas in the whole temperature range, the second from top row shows fraction of hot ($T>10^5$ K) gas, the third -- fraction of warm  $10^3<T<10^5$ K gas, while the lower -- fraction of low-temperature ($T<10^3$ K) phase.  By definition $\int_{-\infty}^{\infty} f_m(z) dz=1$,  $\int_{-\infty}^{\infty} f_{m,h}(z) dz=\mu_h$, $\int_{-\infty}^{\infty} f_{m,w}(z) dz=\mu_w$, $\int_{-\infty}^{\infty} f_{m,c}(z) dz=\mu_c$, and $\mu_h+\mu_w+\mu_c=1$. It is readily seen that most of the gas mass elevates during around 10 to 25 Myr depending on {SN} rate, and then reverses to fall on to disc regardless of continuous SN explosions. At the same time warm and hot gas continues to be elevated at longer times $t>30$ Myr. This conclusion looks consistent with observations of extended hot (X-ray) haloes around edge-on galaxies with a relatively modest SF rate. For instance, observations on XMM-{\it Newton} and {\it Chandra} of an edge-on galaxy NGC 891 reveal presence of hot ($T\sim 0.3$ keV) gas extending up to $z\sim 10$ kpc, which is comparable to the galaxy radius \citep{hodges18}. } 

{An important feature seen in Figure \ref{fig-mass-t} is worth noting: models with low SN and SF rates 
($\nu\leq 2\times 10^{-13}$~yr$^{-1}$~pc$^{-3}$ and $\Sigma_{_{\rm SF}}\simeq 6\times 10^{-3}z_{\ast,100}M_\odot$~yr$^{-1}$~kpc$^{-2}$) reveal a very weak elevation of gas at heights $z<300$ pc. 
It corresponds to a `puffing out' of circulations at low height mentioned above, and the value 
$\nu\leq 2\times 10^{-13}$~yr$^{-1}$~pc$^{-3}$ equivalent to the surface energy input rate 
$\leq 1.4\times 10^{-4}z_{*,100}$~erg~s$^{-1}$~cm$^{-2}$; here the surface SN rate has been defined as a product of the volumetric SN rate by two  stellar scale heights $2z_{\ast}$, see Eq. (\ref{sigm}). For $z_{\ast}=300$ as stated in Sec. \ref{modl} the energy threshold is $\leq 4\times 10^{-4}$~erg~s$^{-1}$~cm$^{-2}$. This value may be treated as the threshold energy input rate necessary for driving 
gas outflows in the regime with disc-wide SN explosions. 
}

{This effect is intimately connected to a long-standing problem of "mass entraiment" into shock driven outflow meaning that hot dilute gas is inefficient to accelerate cold dense clouds \citep[see detailed discussion in][]{scana15,schn17,zhang17}. It is connected with the fact that a high density contrast between dense clouds and dilute hot gas is typically $\chi>100$, and characteristic acceleration time is much longer than the cloud crashing time. In order to overcome this problem \citet{oh18} recently suggested that an additional radiative cooling in a turbulent mixing layer stimulates enhancement of mass loading of the outflowing low density gas. However, for typical parameters of clouds and ambient gas it may take $\sim 30$ Myr which is longer than local starburt events in discs. 
}

\begin{figure*}
\center
\includegraphics[width=14cm]{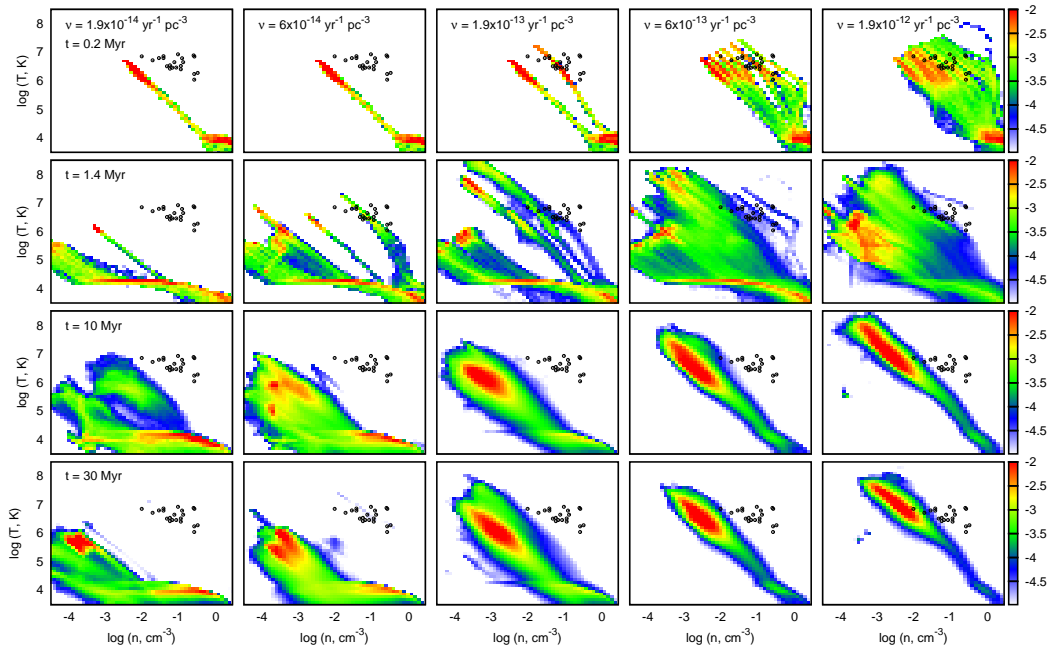}
\caption{
The temperature-density distribution for the models  with $\nu=2\times 10^{-14}$, ~$6\times 10^{-14}$,~$2\times 10^{-13}$, ~$6\times 10^{-13}
$,~$2\times 10^{-12}$ yr$^{-1}$ pc$^{-3}$ from left to right; time {moments} are 0.2, 1.4, 10 and 30 Myr from top to bottom.
The black points depict the observational data for interstellar superbubbles taken from Table~3 of \citet{gupta18}.
}
\label{fig-tn-histo}
\end{figure*}


\section{Observational consequences}

Hot low density gas inside expanding bubbles and outflows can be observed in emission and absorption in far-UV and X-ray ranges.  
Black points in Figure~\ref{fig-tn-histo} depict 'temperature--density' relations from available observational data compiled in 
\citet[][ see their Table 3]{gupta18}. For comparison we show scatter plots connecting temperature and density through the whole 
computational zone for the models with $2\times 10^{-14}$,~$6\times 10^{-14}$,~$2\times 10^{-13}$, ~$6\times 10^{-13}$, 
~$2\times 10^{-12}$ yr$^{-1}$ pc$^{-3}$ at several epochs. The two upper rows correspond to early times: 0.2 and 1.4~Myrs, 
whereas the two lower correspond to the middle time, 10~Myr, and at the final epoch, 30~Myr. A 
noticeable trend is that only young 
dynamical states -- $t\simlt 1.5$ Myr of the ISM with a high SF rate ($\nu\geq 2\times 10^{-13}$ yr~pc$^{-3}$) contain hot and 
relatively dense gas seen observationally. This may arise from the fact that the X-ray emission from dilute hot gas does not contribute much and  therefore is missed in observations compiled in \citet{gupta18}. On the other hand, it is a natural 
consequence of a fast cooling ($t\simlt 3$ Myr) of X-ray gas with density in the range $0.03<n<0.3$ cm$^{-3}$ -- at longer times 
gas temperature falls below the observational level even in the low density range $n\sim 10^{-2}$ cm$^{-3}$ and at high energy injection rate as seen in 3rd and 4th panels in Figure~\ref{fig-tn-histo}. 
It is also clear that lower SN rates produce less hot gas in the whole relevant density range -- the characteristic time between two subsequent 
shock waves impinging a given gas element for $\nu \simlt 2\times 10^{-13}$~yr$^{-1}$ pc$^{-3}$ is longer than the cooling time 
$t\simlt 0.3$ Myr.

\begin{figure*}
\center
\includegraphics[width=14cm]{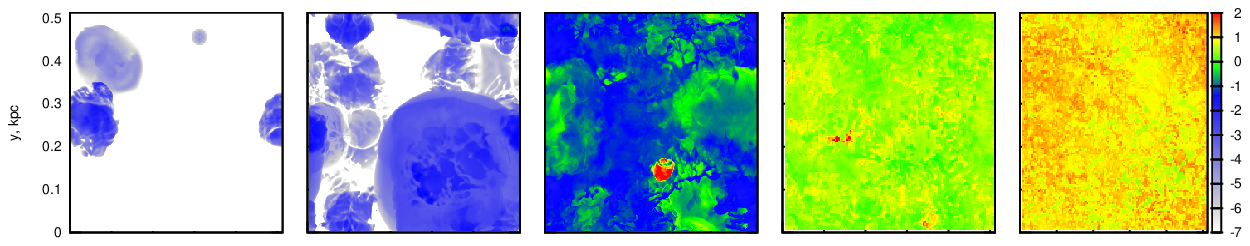}
\includegraphics[width=14cm]{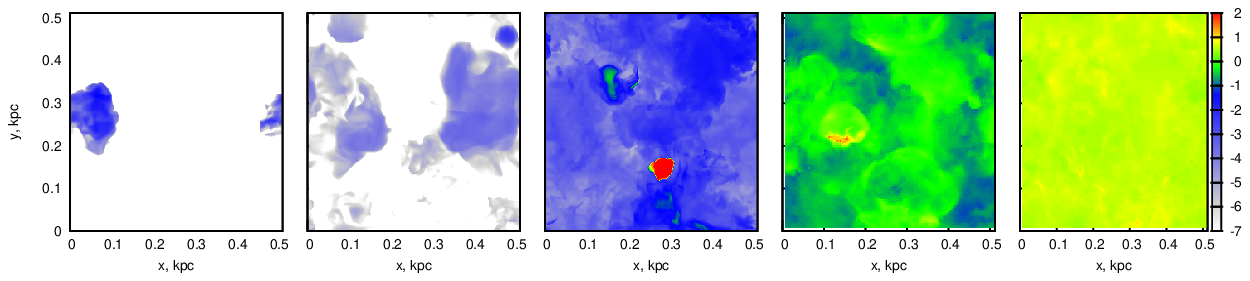}
\caption{
The X-ray surface brightness ($\log(I_X, {\rm keV~cm^{-2}~s^{-1}~str^{-1}})$) maps in the ROSAT bands $0.1\hbox{--}0.3$~keV (upper row) and $1.6-8.3$~keV (lower row) for the models depicted on Fig. \ref{fig-1sn-por-10M} (from left to right) at $10$~Myr.
}
\label{fig-xmaps-10Myr}
\end{figure*}

{
Figure~\ref{fig-xmaps-10Myr} presents the X-ray surface brightness $\log(I_X, {\rm keV cm^{-2} s^{-1} sr^{-1}})$ in the ROSAT bands $0.1-0.3$~keV 
(upper row) and $1.6-8.3$~keV (lower row), for the models depicted on Figure~\ref{fig-1sn-por-10M} (from left to right) at $10$~Myr. {Here we assume that the observer looks along the $z$-direction, in order to determine the maps in $xy$-plane.} {Absorptions by the gas layer on the line-of-sight between the emitting gas and the observer is accounted with the cross-section given by \citet{wilms00}. As the absorption decreases with energy approximately as $\sigma(E)\propto E^{-2.5}$, the high energy ROSAT band map is insensitive to absorption.} For SNe rate 
$\nu \simgt 2\times 10^{-13}n^{4/3}$~yr$^{-1}$ pc$^{-3}$ with $n$ being mean ambient density, 
the surface brightness in the low-energy band at $t\simgt 10$~Myr remains greater than 
$10^{-2}~{\rm keV~cm^{-2}~s^{-1}~str^{-1}}$ inside the simulated area. This means that such SNe rate produces a {rather smooth steady 
state X-ray field -- the absorption does not significantly change the brightness distribution, it shows only a difference within factor of 2 on small 
scales.} 
For lower rates though, the distribution of the surface X-ray brightness is patchy -- only separate regions of the ISM disc emit in this band within 
one cooling time and then disappear, making the soft X-ray surface pattern intermittent in space and time. In the high-energy a steady state 
X-ray field can be supported by SNe rate $\nu \simgt 6\times 10^{-13}n^{4/3}$~yr$^{-1}$ pc$^{-3}$. 
}

\begin{figure}
\center
\includegraphics[width=7cm]{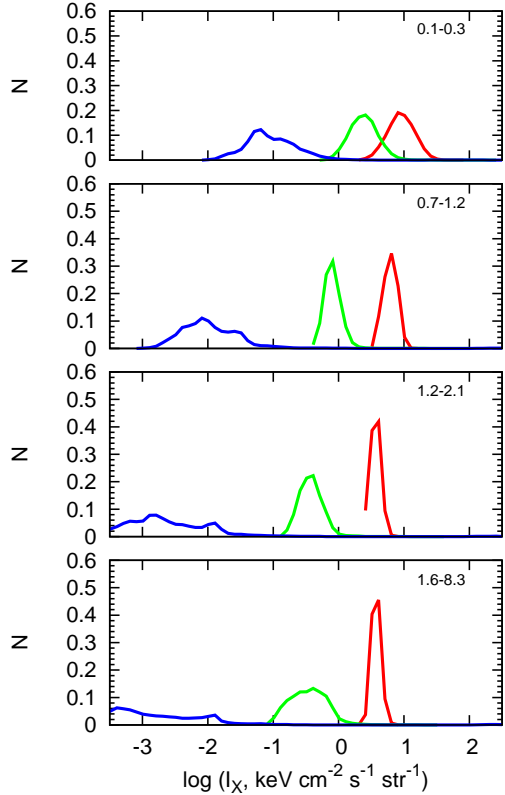}
\caption{
Distribution functions (histograms) of X-ray surface brighness in the ROSAT bands (0.1-0-3, 0.7-1.2, 1.2-2.1, 1.6-8.3~keV from top to bottom) for 
the models  with~$2\times 10^{-13}$~yr$^{-1}$ pc$^{-3}$ (blue), ~$6\times 10^{-13}$ (green),~ $\nu=2\times 10^{-12}$ (red); {stable smooth X-ray fields with  narrow distributions are kept at 10~Myr only for models with $\nu>2\times 10^{-13}$~yr$^{-1}$ pc$^{-3}$, for lower SN rates the  distributions are wide and nearly flat} (see Figure~\ref{fig-xmaps-10Myr}). 
}
\label{fig-xhisto-10Myr}
\end{figure}

Figure~\ref{fig-xhisto-10Myr} presents the distribution functions of X-ray surface brighness {-- the  probability to achieve the brightness in a given range,} in the ROSAT bands for the models with $\nu=2\times 10^{-13}$, ~$6\times 10^{-13}$,~$2\times 10^{-12}$~yr$^{-1}$ pc$^{-3}$ for which a steady state X-ray field remains at 10~Myr (see Figure~\ref{fig-xmaps-10Myr}). One can see a shift of the distributions towards higher energy bands as predicted  in \citep{vsn17}. For the low-energy range, 0.1-0.3~keV, distributions for various SNe rates are similar with a single sharp peak. One may expect a sort of relation between the peak value of surface brightness and SNe rate.

\begin{figure}
\center
\includegraphics[width=7cm]{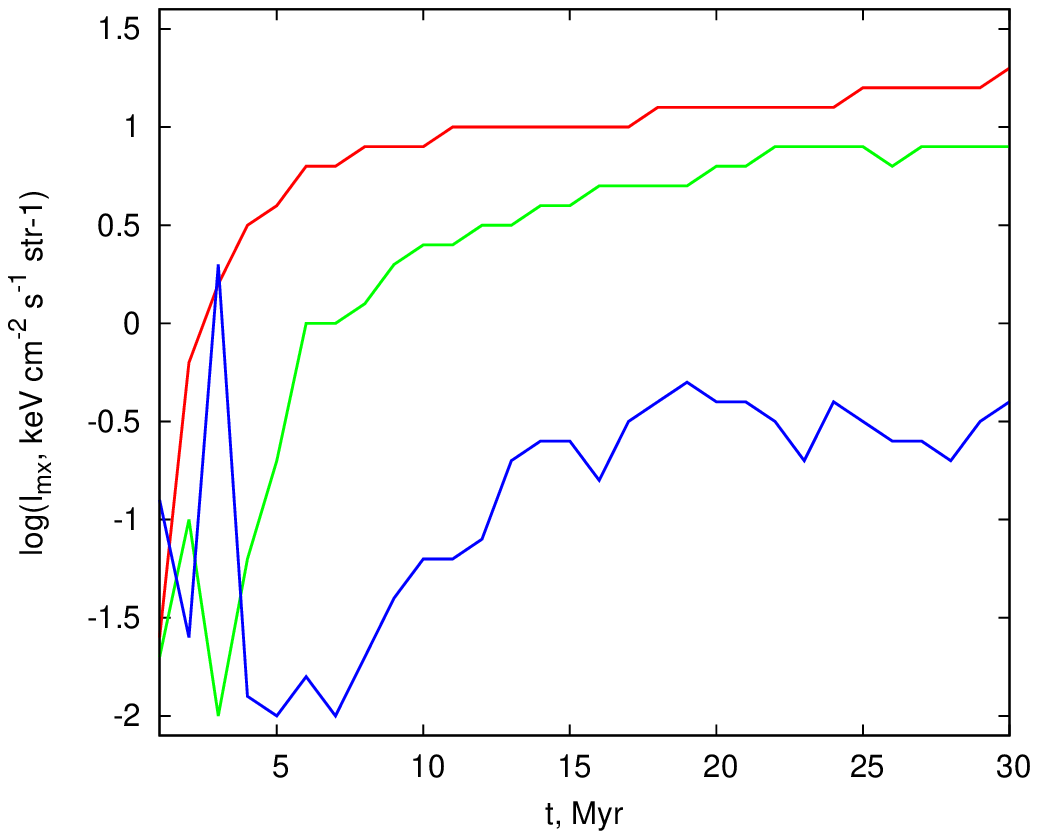}
\includegraphics[width=7cm]{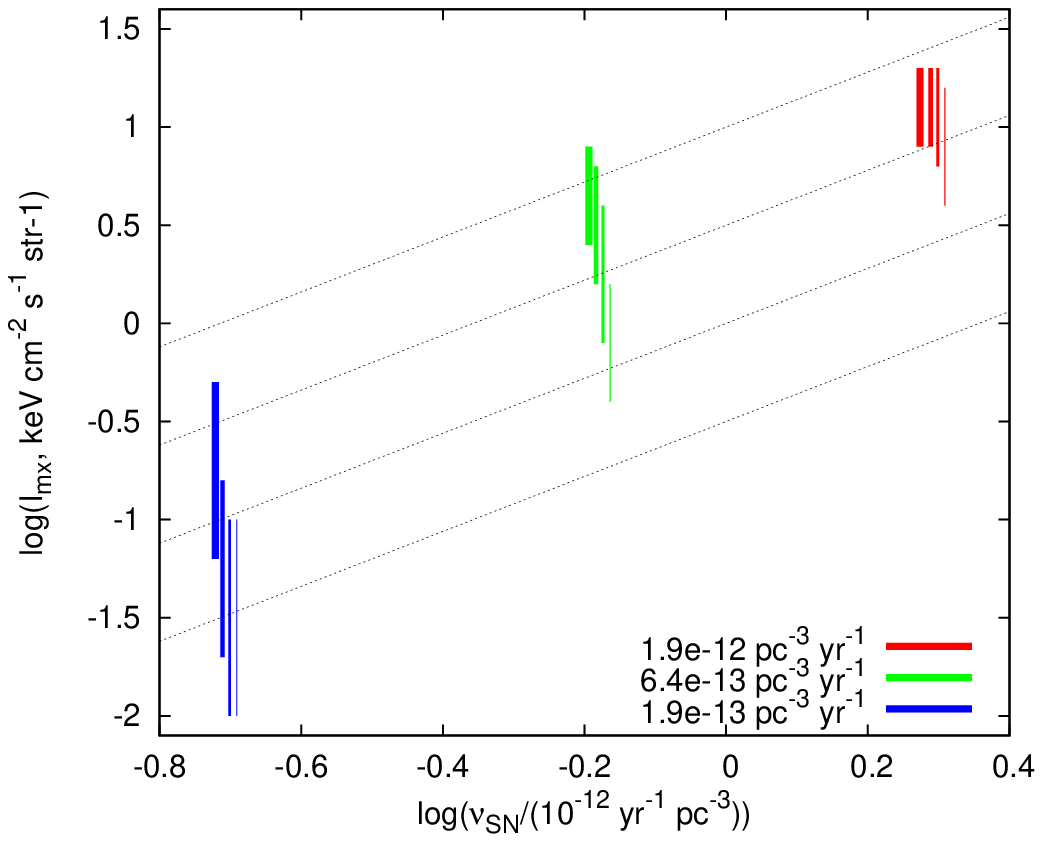}
\caption{
{\it Top panel.} The evolution of the peak value of X-ray surface brightness as shown in the distribution function for 0.1-0.3~keV band on upper panel of Figure~\ref{fig-xmaps-10Myr}. {\it Bottom panel.} The dependence "peak of X-ray brightness -- SN rate". The color vertical lines correspond to the dispersal of the peak during the evoultion from 10 to 30~Myr: from left to right they correspond to the bands 0.1-0.3, 0.7-1.2, 1.2-2.1, 1.6-8.3~keV, for the sake of clearity they have different thickness and shifted by $\Delta\log\nu_{\rm SN}=0.01$. The dotted lines show  linear functions $\log(I_{\rm mx}\simeq 1.4\log\nu_{\rm SN}-(0.2+0.5k)$, the value of $k$ changes from soft to hard bands.
}
\label{fig-xmax-snrate}
\end{figure}

In order to check this assumption we follow the evolution of the peak  value for three values of the SNe rate able to keep a steady state X-ray field within the first few million years. In top panel of Figure~\ref{fig-xmax-snrate} one can observe that the magnitude of the peak brightness $I_{mx}$ saturates with time for 0.1-0.3~keV band. The bottom panel of Figure~\ref{fig-xmax-snrate} shows the relation "peak X-ray brightness -- SN rate" for different bands, the ambient gas density $n=1$~cm$^{-3}$ is assumed; thin dotted lines $\log(I_{\rm mx}\simeq\log\nu_{\rm SN}-(0.2+0.5k)$, where $k=0,~1,~2,~3$ count the X-ray bands from the softest $E=0.1\hbox{--}0.3$ keV to the hardest $E=1.6\hbox{--}8.3$ keV approximate the data. Colored vertical lines show the spread of the X-ray brightness peak during the evolution from 10 to 30 Myr; for different bands lines have different thickness and are shifted right by $\Delta\log\nu_{\rm SN}=0.01$.  {This allows us to independently estimate the SN rate from X-ray observations.}

\begin{figure}
\center
\includegraphics[width=7cm]{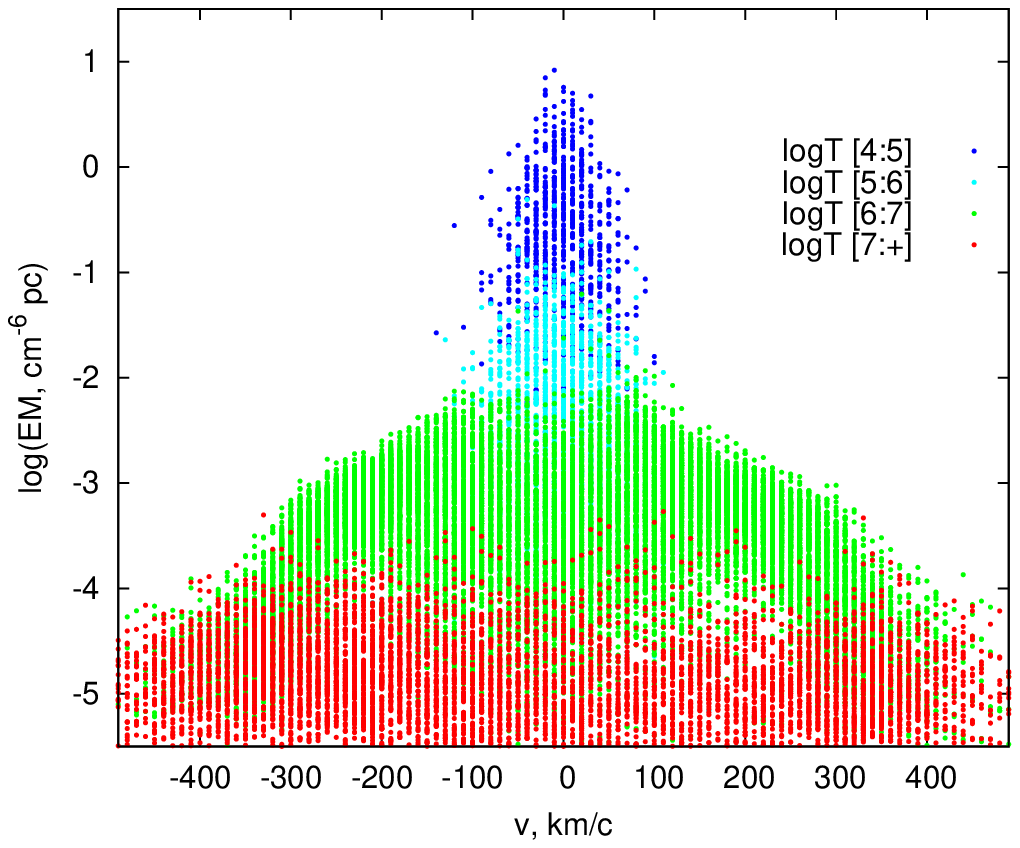}
\includegraphics[width=7cm]{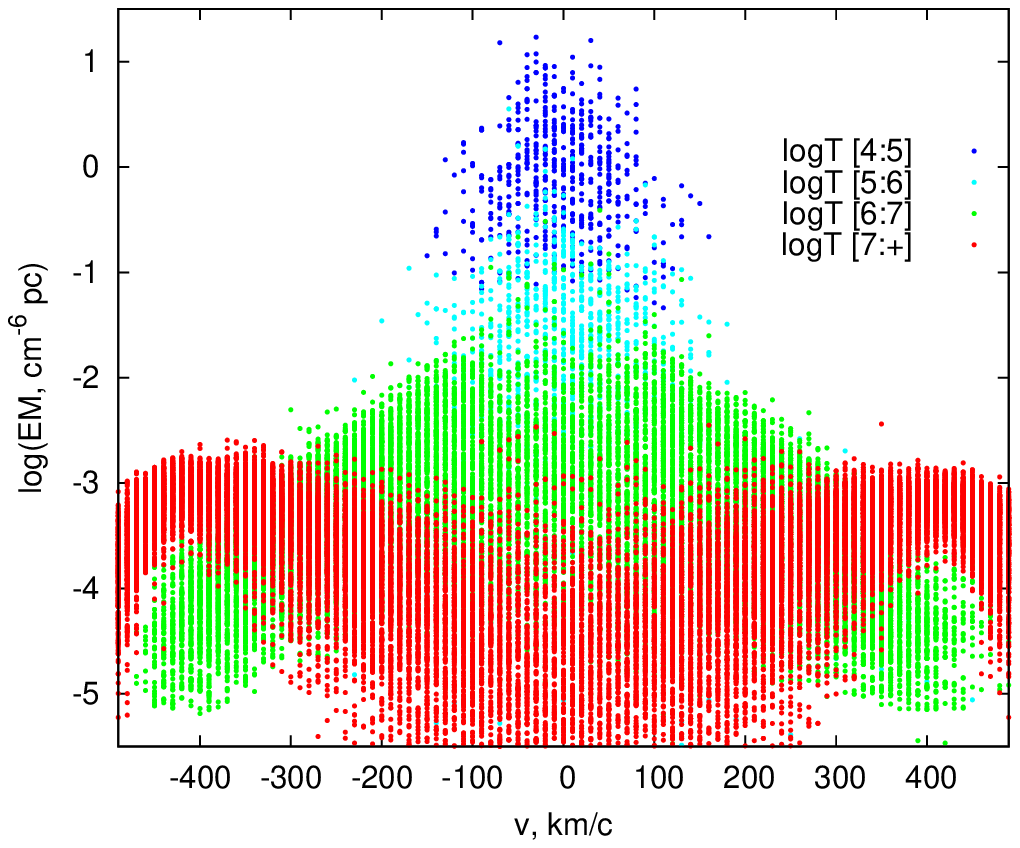}
\caption{
Dependence of emission measure versus velocity along the sight-line perpendicular to the disc for different temperature bins as shown in color legends; upper panel for SN rate $\nu = 6\times 10^{-13}$~yr$^{-1}$~pc$^{-3}$, lower panel for $\nu = 2\times 10^{-12}$~yr$^{-1}$~pc$^{-3}$.
}
\label{fig-intensity-velocity}
\end{figure}

{In order to obtain spectral features of galactic outflows widely distributed through over interstellar discs we calculate the dependence of emission measure in different temperature bins versus velocity along several lines of sight perpendicular to the disc. These values  calculated in numerical cells within the velocity $[v,v+\Delta v]$ are believed to reflect both optical and X-ray emissivity of outflows (e.g. in lines of metal ions like CIV, OVI, OVII, NeVIII, NeIX, MgX etc.); the velocity ranges from $-500$ km s$^{-1}$ to $+500$ km s$^{-1}$. {Such disc-wide outflows appear to be relatively quiescent as compared to violent explosive events in galaxies with a central starburts, as for instance, in M82 galaxy. Their spectral features are also less pronounced than in the case of strong winds from galactic centres}. 

Figure \ref{fig-intensity-velocity} shows scatter plots for the emission measure along 300 lines of sight in $z$-direction in the model with $\nu=6\times 10^{-13}$~yr$^{-1}$~pc$^{-3}$ ($\Sigma_{_{\rm SF}}\simeq 0.02~M_\odot$~yr$^{-1}$~kpc$^{-2}$, upper panel) and $\nu=2\times 10^{-12}$~yr$^{-1}$~pc$^{-3}$ ($\Sigma_{_{\rm SF}}\simeq 0.06~M_\odot$~yr$^{-1}$~kpc$^{-2}$, lower panel) for $t=30$ Myr. It is seen that the EM-velocity profiles  reveal a distinct two-peak shape corresponding to outflowing gas at $v=\pm 100$ km s$^{-1}$ for lower SN rate and at $v=\pm 200$ km s$^{-1}$ for higher SN rate for the temperature bin $T=10^6\hbox{--}10^7$ K, while the EM value at higher temperature bin $T\geq 10^7$ K shows the peaks at $v=\pm 250$ km s$^{-1}$ for smaller SN rate and $v=\pm 400$ km s$^{-1}$ for higher SN rate. Note that there is no two-peak velocity distribution for SN rate $\nu\simlt 2\times 10^{-13}$~yr$^{-1}$~pc$^{-3}$, because there is no outflow signature for such SN rate. In this case the velocity distribution has one peak around zero with dispersion lower than 50~km~s$^{-1}$. Therefore measurements of such profiles can provide an estimate of hot gas density and the correspponding mass outflow rate. It is worth noting, that the peak velocities are factor of $\sqrt{A}$ higher than the thermal Doppler line width, where $A$ is atomic number of the ion, and thus macroscopic outflow motion seems to be easily recognizable. Moreover, an additional feature for discrimination of thermal line broadening and its shift due to macroscopic motion is the non-gaussian line profile for the latter case.    
}

\section{Conclusions}

\noindent
We have studied how isolated supernova remnants spread through over the disc merge into collective bubbles depending on the SN rate and 
eject interstellar gas into haloes. Our results are summarized as follows:  

\begin{itemize}
 \item In the initial episodes of SNe activity during $\sim 1$ Myr gas cools radiatively, fragments and forms dense clumps. On next stages 
 two flows do form: high velocity ($|v|\geq 70$ km s$^{-1}$) hot diffuse gas along with low velocity ($|v|\leq 70$ km s$^{-1}$) dense gas 
 moving outward, and counter flowing low velocity ($|v|\leq 70$ km s$^{-1}$) dense clumps moving inward. This conclusion is consistent with 
 estimates inferred from numerical simulations by \cite{kim18} within the vertical scales of our models, though the mass loading factor -- the ratio of mass outflow rate to the SF rate, for hot gas is an order of magnitude lower $\sim 0.01nz_{100}$.
 
 \item {Within the regime of disc-wide SN explosions the threshold energy input rate for driving hot gas outflows can be extimated as $\sim 4\times 10^{-4}$~erg~s$^{-1}$~cm$^{-2}$ for stellar scale height $z_{\ast}=300$ pc.}
 
 \item After $t\simgt 10$ Myr outward motion weakens under gravity and eventually only a small fraction $\sim 3\times 10^{-3}$ of cold gas remains 
 at heights $\simgt 0.5$ kpc even for the rates $\sim 2\times 10^{-12}$ yr$^{-1}$ pc$^{-3}$. Hot {and warm} gas stay at heights $\sim 1$ kpc on a longer time scale $>30$ Myr. {This results is consistent with observations of X-ray coronae around edge-on galaxies with a modest star formation rate $\nu\sim 2\times 10^{-13}$ yr$^{-1}$ pc$^{-3}$.  For lower SN rates $\nu<2\times 10^{-14}$ yr$^{-1}$ pc$^{-3}$ its mass fraction falls though below $\sim 10^{-5}$.} 
 
 \item The interrelation between temperature and density in gas confined in bubbles nearly corresponds to an isobaric conditions (though with a high spread). Observational X-ray data \citep[as compiled in][]{gupta18} likely correspond to young gas ($t\simlt 1.5$ Myr) excited by SN explosions with a relatively high rate $\nu \simgt 2\times 10^{-13}$~yr$^{-1}$ pc$^{-3}$. 
 
 \item When seen face-on interstellar discs reveal either steady state or intermittent in space and time distribution of surface brightness through over the disc. {High SN rate models with $\nu \simgt 2\times 10^{-13}$~yr$^{-1}$ pc$^{-3}$ show one-peak brightness distribution function $F(I_{\rm X})$ in different energy bands with the peak brightness nearly linearly growing with} SN rate.   
 
 \item Emission line profiles of highly ionized ions present in hot ($T\geq 10^6$ K) outflowing gas can serve as tracers of the outflow and allows to estimate hot gas density and the outflow mass rate.

\end{itemize}

\section{Acknowledgements}

\noindent

YS thanks V. Dwarkadas for discusssion. EV is grateful to the Ministry for Education and Science of the Russian Federation (grant 3.858.2017/4.6). The simulations have been supported by the Russian Scientific Foundation (grant 14-50-00043). YS is partially supported by the RFBR (grant 17-52-45053), by the project 01-2018 ``New Scientific Groups LPI'', and by the Program of the Presidium of RAS (project code 28). YS aknowledges support from Raman Research Institute where this work has been partly conducted.


\end{document}